\def\be{\begin{equation}}
\def\ee{\end{equation}}
\def\ba{\begin{array}}
\def\ea{\end{array}}

\documentclass[aps,amsmath,amssymb,amsfonts,pra]{revtex4}
\usepackage{graphicx}
\usepackage{epstopdf}
\usepackage{multirow}
 \usepackage [latin1]{inputenc}
\def\qed{\leavevmode\unskip\penalty9999 \hbox{}\nobreak\hfill
     \quad\hbox{\leavevmode  \hbox to.77778em{%
               \hfil\vrule   \vbox to.675em%
               {\hrule width.6em\vfil\hrule}\vrule\hfil}}
     \par\vskip3pt}

\newtheorem{theorem}{Theorem}
\newtheorem{corollary}{Corollary}

\newtheorem{proposition}{Proposition}
\begin{document}

\title{On fully entangled fraction of arbitrary $d\otimes d$ quantum states}

\author{Xue-Na Zhu$^{1}$}
\thanks{jing\_feng1986@126.com}
\author{Gui Bao$^{1}$}
\author{Ming Li$^{2}$}
\author{Ming-Jing Zhao$^{3}$}
\author{Shao-Ming Fei$^{4}$}
\thanks{feishm@cnu.edu.cn}
\affiliation{$^1$School of Mathematics and Statistics Science, Ludong University, Yantai 264025, China\\
$^2$College of the Science, China University of Petroleum, Qingdao 266580, China\\
$^3$ School of Science, Beijing Information Science and Technology University, Beijing 100048, China\\
$^4$School of Mathematical Sciences, Capital Normal University, Beijing 100048, China}

\begin{abstract}
We study the fully entangled fraction of quantum states based on the Bloch representation of density matrices. Analytical upper bounds on the fully entangled fraction are obtained for arbitrary $d\otimes d$ bipartite systems. The fully entangled fractions for classes of $d\otimes d$ quantum states are analytically derived. Detailed examples are given to illustrate the advantages of our
results.
\end{abstract}

\maketitle

\section{Introduction}

Quantum teleportation \cite{t1,t2} plays an important role in quantum information processing, demonstrating the powerful applications of quantum correlations in transmitting an unknown quantum state from the sender to a spatially separated receiver.
It is shown that the optimal fidelity of teleportation is given by the fully entangled fraction (FEF) of the quantum resource shared by the sender and the receiver \cite{tf}, rather than the entanglement of the quantum resource.
The quantity  FEF plays also essential roles in many other quantum information processing such as quantum dense coding
\cite{cd1,cd2}, entanglement swapping \cite{es1,es2} and quantum cryptography (Bell inequalities) \cite{B2,B3}. Thus an analytic estimation of FEF is of great importance.

Denote $H_A$ and $H_B$ the vector spaces with dimension $d$.
The FEF of a density matrix $\rho\in H_A\otimes H_B$ is defined by \cite{tf,PRA601888}
\begin{equation}\label{FEF}
F(\rho)=\max_{U}\langle\phi_{+}|U^{\dag}\otimes I\rho U\otimes I|\phi_{+}\rangle,
\end{equation}
where $U$ (resp. $I$) are unitary (resp. identity) matrices, and $|\phi_{+}\rangle=\frac{1}{\sqrt{d}}\sum_{s=0}^{d-1}|ss\rangle$ is the maximally entangled state.

Considerable efforts have been made towards the evaluation of FEF both analytically and numerically
for some mixed states \cite{2FEF,dd4,HM1,HM2,up,huang}.
Let $\lambda_i$, $i=1,2,...,d^2-1$, be the generators of $SU(d)$ given by $\{\omega_m,u_{lk},v_{lk}\}$ with
$\omega_m=\sqrt{\frac{2}{(m+1)(m+2)}}\left(\sum_{t=0}^{m}|t\rangle\langle t|-(m+1)|m+1\rangle\langle m+1|\right)$, $u_{lk}=|l\rangle\langle k|+|k\rangle\langle l|$ and $v_{lk}=-\mathfrak{i}(|l\rangle\langle k|-|k\rangle\langle l|)$, $0\leq m\leq d-2$, $0\leq l< k\leq d-1$, where $\mathfrak{i}^2=-1$.
A bipartite state $\rho\in H_A\otimes H_B$ can be written in the following Bloch representation \cite{JR},
\begin{equation}\label{rho}
\rho=\frac{1}{d^2}\left(I\otimes I+\sum_{i=1}^{d^2-1}r_i\lambda_i\otimes I+\sum_{j=1}^{d^2-1}s_j I\otimes\lambda_j+\sum_{i,j=1}^{d^2-1}t_{ij}\lambda_i\otimes \lambda_j\right),
\end{equation}
where $r_i=\frac{d}{2}Tr(\rho\lambda_i\otimes I)$, $s_i=\frac{d}{2}Tr(\rho I\otimes\lambda_i)$ and $t_{ij}=\frac{d^2}{4}Tr(\rho\lambda_i\otimes \lambda_j)$.

In Ref.\cite{2FEF}, an elegant formula of FEF for two-qubit systems is analytically derived by using
the method of Lagrange multiplier. Denote $T=[t_{ij}]$ the matrix given by the entries $t_{ij}$ in (\ref{rho}) and $||T||_{KF}=Tr(\sqrt{T^{\dagger}T})$ the Ky-Fan norm of $T$. The FEF of
any two-qubit state $\rho$ is given by \cite{HM1,HM2},
\begin{equation*}
F(\rho)=\frac{1+||T||_{KF}}{4}.
\end{equation*}
For general high-dimensional quantum states, the analytical computation of FEF remains formidable. In Ref.\cite{dd4}, the fully entangled fractions of the isotropic states and Werner states are analytically computed. The
upper bound of the FEF has been estimated by Refs.\cite{up,huang}.
The upper bound derived in Ref.\cite{up} says that for any $d\otimes d$ quantum states $\rho$, the FEF
$F(\rho)$ satisfies the following inequality,
\begin{equation*}
F(\rho)\leq\frac{1}{d^2}+4||M^{T}(\rho)M(|\phi_{+}\rangle\langle\phi_{+}|)||_{KF},
\end{equation*}
where $M^{T}$ stands for the transpose of $M$, and
$M(\rho)$ denote the correlation matrix with entries $m_{ij}=\frac{1}{d^2}t_{ij}$.

In this paper, we study the FEF based on the Bloch representation of density matrices in $d\otimes d$ systems. First, we simplify the definition (\ref{FEF}) to a simpler expression. Based on this expression, we present the upper and lower bounds of FEF. Furthermore, the FEF of some special states are analytically derived.

\section{An alternative expression of the fully entangled fraction}

For convenience, denote $\Omega_1=\{1,...,d-1\}$, $\Omega_2=\{d,...,\frac{1}{2}(d-1)(d+2)\}$, $\Omega_3=\{\frac{1}{2}d(d+1),...,d^2-1\}$ and $\Omega=\Omega_1\cup\Omega_2\cup\Omega_3$.
Without loss of generality, we suppose that
\begin{equation*}\label{y}
\lambda_i=\left\{
\begin{aligned}
&~~~ \omega_{i-1},~~~ ~~~~ i\in\Omega_1;\\
&~~~u_{lk},~~~~~~~~~i\in\Omega_2~and~0\leq l<k\leq d-1;\\
&~~~v_{lk},~~~~~~~~~ i\in\Omega_3~and~0\leq l<k\leq d-1.
\end{aligned}
\right.
\end{equation*}
It can be straightforwardly verified that
\begin{equation}\label{y1}
\langle s|\lambda_i|s\rangle=\left\{
\begin{aligned}
& \sqrt{\frac{2}{i(i+1)}},~~~~~ ~~~~ i\in\Omega_1~and~~0\leq s\leq i-1 ;\\
&-i\sqrt{\frac{2}{i(i+1)}},~~~~~~~i\in\Omega_1~and~~ s=i;\\
&~~~~~~0,~~~~~~~~~~~~~~~~~~i\in\Omega_1~and~ s\geq i+1;\\
&~~~~~~ 0,~~~ ~~~~~~~~~~~~~~~ i\in\Omega_2\cup\Omega_3,
\end{aligned}
\right.
\end{equation}
and
\begin{equation}\label{y2}
\langle t|\lambda_i|s\rangle=\left\{
\begin{aligned}
& 0,~~~~~ ~~~~~i\in\Omega_1 ;\\
&1,~~~~~ ~~~~ ~i\in\Omega_2,~\{s,t\}=\{l,k\}~~~and~0\leq l<k\leq d-1;\\
&\mathfrak{i} ,~~ ~~~~~i\in\Omega_3~and~0\leq s=l<k=t\leq d-1;\\
&-\mathfrak{i},~~~~~ ~~ ~~~~i\in\Omega_3~and~0\leq t=l<k=s\leq d-1,
\end{aligned}
\right.
\end{equation}
where $s\not=t$.

\begin{proposition}\label{p1}
The fully entangled fraction of a state $\rho\in H_A\otimes H_B$ can be expressed as
 \begin{eqnarray}\label{F}
F(\rho)&=&\frac{1}{d^2}+\frac{1}{d^3}\max_{U}\left(\sum_{i\in\Omega,j\in\Omega_1\cup\Omega_2}
t_{ij}Tr(U^{\dag}\lambda_{i}U\lambda_j)-\sum_{i\in\Omega,j\in\Omega_3}
t_{ij}Tr(U^{\dag}\lambda_{i}U\lambda_j)\right).
\end{eqnarray}
\end{proposition}

{\sf [Proof]}~
 Combining  the definition (\ref{FEF}) and  the Bloch representation of $\rho$,  we have
\begin{eqnarray*}
&&\langle\phi_{+}|U^{\dag}\otimes I\rho U\otimes I|\phi_{+}\rangle
=\frac{1}{d^2}\big(1+\sum_{i\in\Omega}r_i\langle\phi_{+}|U^{\dag}\lambda_{i}U\otimes I|\phi_{+}\rangle+
\sum_{j\in\Omega}s_j \langle\phi_{+}|I\otimes \lambda_{j}|\phi_{+}\rangle+\sum_{i,j\in\Omega}t_{ij}\langle\phi_{+}|U^{\dag}\lambda_{i}U\otimes \lambda_{j}|\phi_{+}\rangle\big).
\end{eqnarray*}
Denote $R(U)=\sum_{i\in\Omega}r_i\langle\phi_{+}|U^{\dag}\lambda_{i}U\otimes I|\phi_{+}\rangle$,
$S(U)=\sum_{j\in\Omega}s_j \langle\phi_{+}|I\otimes \lambda_{j}|\phi_{+}\rangle$ and
$T(U)=\sum_{i,j\in\Omega}t_{ij}\langle\phi_{+}|U^{\dag}\lambda_{i}U\otimes \lambda_{j}|\phi_{+}\rangle.$
Taking into account the property that $Tr((M\otimes I)\rho)=Tr(M\rho_A)$,  we obtain
\begin{eqnarray*}
R(U)&=&\sum_{i\in\Omega}r_iTr[(U^{\dag}\lambda_{i}U\otimes I)\Phi^{+}]\\\nonumber
&=&\sum_{i\in\Omega}r_iTr[U^{\dag}\lambda_{i}U\Phi^{+}_{A}]\\\nonumber
&=&\frac{1}{d}\sum_{i\in\Omega}r_iTr[U^{\dag}\lambda_{i}U]=0,
\end{eqnarray*}
where $\Phi^{+}=|\phi_{+}\rangle\langle\phi_{+}|$ and
$\Phi^{+}_{A}=Tr_{B}(\Phi^{+})=\frac{1}{d}I.$
Similarly, we have
$S(U)=0$.

On the other hand,
\begin{eqnarray*}\label{FXY}
T(U)
&=&\sum_{i,j\in\Omega}t_{ij}Tr[U^{\dag}\lambda_{i}U\otimes \lambda_{j}\Phi^{+}]\\\nonumber
&=&\frac{1}{d}\sum_{i,j\in\Omega}t_{ij}\sum_{s,t=0}^{d-1}
Tr[(U^{\dag}\lambda_{i}U\otimes \lambda_{j})(|s\rangle\langle t|\otimes|s\rangle\langle t|)]\\\nonumber
&=&\frac{1}{d}\sum_{i,j\in\Omega}\sum_{s,t=0}^{d-1}t_{ij}\langle t|U^{\dag}\lambda_{i}U|s\rangle
\langle t|\lambda_{j}|s\rangle\\\nonumber
&=&\frac{1}{d}\sum_{i,j\in\Omega}\sum_{s=0}^{d-1}t_{ij}\langle s|U^{\dag}\lambda_{i}U|s\rangle
\langle s|\lambda_{j}|s\rangle+\frac{1}{d}\sum_{i,j\in\Omega}\sum_{s\neq t}t_{ij}\langle t|U^{\dag}\lambda_{i}U|s\rangle
\langle t|\lambda_{j}|s\rangle\\\nonumber
&\equiv&\frac{1}{d}(T_1(U)+T_2(U)),
\end{eqnarray*}
where $T_1(U)=\sum_{i,j\in\Omega}\sum_{s=0}^{d-1}t_{ij}\langle s|U^{\dag}\lambda_{i}U|s\rangle
\langle s|\lambda_{j}|s\rangle$ and $T_2(U)=\sum_{i,j\in\Omega}\sum_{s\neq t}t_{ij}\langle t|U^{\dag}\lambda_{i}U|s\rangle
\langle t|\lambda_{j}|s\rangle$. Eq.(\ref{y1}) implies that
(A1) if $j\in \Omega_1$, then $\langle s|\lambda_{j}|s\rangle=\sqrt{\frac{2}{j(j+1)}}$ for $s=0,...,j-1$,
$\langle s|\lambda_{j}|s\rangle=-j\sqrt{\frac{2}{j(j+1)}}$ for $s=j$,
$\langle s|\lambda_{j}|s\rangle=0$ for $s=j+1,...,d-1$; (A2) if  $j\in \Omega_2\cup\Omega_3$, then $\langle s|\lambda_{j}|s\rangle=0$ for all $s=0,...,d-1$.
By virtue of (A1) and (A2)  we have
\begin{eqnarray}\label{XU}
T_1(U)
&=&\sum_{i\in\Omega}\sum_{j\in\Omega_1}\sum_{s=0}^{d-1}t_{ij}\langle s|U^{\dag}\lambda_{i}U|s\rangle
\langle s|\lambda_{j}|s\rangle\\\nonumber
&=&\sum_{i\in\Omega}\sum_{j\in\Omega_1}t_{ij}\big[\langle 0|U^{\dag}\lambda_{i}U|0\rangle
\langle 0|\lambda_{j}|0\rangle+...+\langle j-1|U^{\dag}\lambda_{i}U|j-1\rangle
\langle j-1|\lambda_{j}|j-1\rangle+\langle j|U^{\dag}\lambda_{i}U|j\rangle
\langle j|\lambda_{j}|j\rangle\big]\\\nonumber
&=&\sum_{i\in\Omega}\sum_{j\in\Omega_1}t_{ij}\big[\sqrt{\frac{2}{j(j+1)}}(\langle 0|U^{\dag}\lambda_{i}U|0\rangle+...+\langle j-1|U^{\dag}\lambda_{i}U|j-1\rangle)-j\sqrt{\frac{2}{j(j+1)}}\langle j|U^{\dag}\lambda_{i}U|j\rangle\big]\\\nonumber
&=&\sum_{i\in\Omega}\sum_{j\in\Omega_1}t_{ij}\sqrt{\frac{2}{j(j+1)}}Tr\big[ U^{\dag}\lambda_{i}U(|0\rangle\langle0|+...+|j-1\rangle\langle j-1|-j|j\rangle\langle j|) \big]\\\nonumber
&=&\sum_{i\in\Omega}\sum_{j\in\Omega_1}t_{ij}Tr(U^{\dag}\lambda_{i}U\lambda_j),
\end{eqnarray}
where the last equation is due to  $\lambda_j=\omega_{j-1}$ for $j\in\Omega_1$.

Note that,  when $j\in\Omega_1$ and $s\not=t$, we have $\langle s|\lambda_{j}|t\rangle=0$ based on Eq.(\ref{y2}).
It remains to prove that
\begin{eqnarray*}\label{tij2}
T_2(U)&=&\sum_{i\in\Omega}\sum_{j\in\Omega_2\cup\Omega_3}\sum_{s\neq t}t_{ij}\langle t|U^{\dag}\lambda_{i}U|s\rangle
\langle t|\lambda_{j}|s\rangle\\
&=&\sum_{i\in\Omega}\sum_{j\in\Omega_2}\sum_{s\neq t}t_{ij}\langle t|U^{\dag}\lambda_{i}U|s\rangle
\langle t|\lambda_{j}|s\rangle+\sum_{i\in\Omega}\sum_{j\in\Omega_3}\sum_{s\neq t}t_{ij}\langle t|U^{\dag}\lambda_{i}U|s\rangle\langle t|\lambda_{j}|s\rangle\\
&\equiv&Y_{21}(U)+Y_{22}(U).
\end{eqnarray*}

(B1) When $j\in\Omega_2$,
 since $\lambda_j=\mu_{lk}$, $0\leq l<k\leq d-1$,  we have
$\langle l|\lambda_{j}|k\rangle=\langle k|\lambda_{j}|l\rangle=1$ and
 $\langle t|\lambda_{j}|s\rangle=\langle s|\lambda_{j}|t\rangle=0$ for all others $s\not=t$.
Hence
 \begin{eqnarray}\label{Y1}
 T_{21}(U)
&=&\sum_{i\in\Omega}\sum_{j\in\Omega_2}t_{ij}\big(\langle k|U^{\dag}\lambda_{i}U|l\rangle
+\langle l|U^{\dag}\lambda_{i}U|k\rangle\big)\\\nonumber
&=&\sum_{i\in\Omega}\sum_{j\in\Omega_2}t_{ij}Tr(U^{\dag}\lambda_{i}U\lambda_j).
\end{eqnarray}

(B2) When $j\in\Omega_3$,
 since $\lambda_j=\nu_{lk}$, $0\leq l<k\leq d-1$, we have
$\langle k|\lambda_{j}|l\rangle=-\langle l|\lambda_{j}|k\rangle=\mathfrak{i}$ and
 $\langle t|\lambda_{j}|s\rangle=\langle s|\lambda_{j}|t\rangle=0$ for all others $s\not=t$.
 Hence
\begin{eqnarray}\label{Y2}
 T_{22}(U)
&=&\sum_{i\in\Omega}\sum_{j\in\Omega_3}t_{ij}\big(\mathfrak{i}\langle k|U^{\dag}\lambda_{i}U|l\rangle
-\mathfrak{i}\langle l|U^{\dag}\lambda_{i}U|k\rangle\big)\\\nonumber
&=&\sum_{i\in\Omega}\sum_{j\in\Omega_3}\left(-t_{ij}Tr(U^{\dag}\lambda_{i}U\lambda_j)\right).
\end{eqnarray}
According to Eqs. (\ref{XU}), (\ref{Y1}) and (\ref{Y2}), we obtain the expression (\ref{F}) of the fully entangled fraction.
$\Box$

\section{ Bounds on the fully entangled fraction}
To study the bounds of FEF based on Eq.(\ref{F}) for arbitrary $d\otimes d$ quantum states $\rho$,
we denote $\Delta(U,i,j)\equiv Tr(U^{\dag}\lambda_{i}U\lambda_j)$ and define
\begin{equation*}
T_1=\sum_{ i, j\in\Omega_1}2(ij+\min\{i,j\})\sqrt{\frac{1}{ij(i+1)(j+1)}}|t_{ij}|,
\end{equation*}
\begin{equation*}
T_2=\sum_{i\in\Omega_1}\sum_{j\in\Omega_2\cup\Omega_3}\sqrt{\frac{2(i+1)}{i}}|t_{ij}|,
\end{equation*}
\begin{equation*}
T_3=\sum_{i\in\Omega_2\cup\Omega_3}\sum_{j\in\Omega_1}\sqrt{\frac{2(j+1)}{j}}|t_{ij}|,
\end{equation*}
and
\begin{equation*}
T_4=2\sum_{i,j\in\Omega_2\cup\Omega_3}|t_{ij}|.
\end{equation*}

\begin{theorem}\label{TH1}
For the quantum state $\rho\in H_A\otimes H_B$, we have
\begin{equation}\label{FUpp}
F(\rho)\leq \frac{1}{d^2}+\frac{1}{d^3}(\sum_{b=1}^{4}T_b).
\end{equation}
\end{theorem}

{\sf [Proof]} It is obvious that $U^{\dag}\lambda_{i}U$ and $\lambda_i$ have  the same  eigenvalues.
(C1) When $i\in\Omega_1$, the  eigenvalues are $\lambda^m_i=\sqrt{\frac{2}{i(i+1)}}$ for $m=1,...,i$, $\lambda^m_i=0$ for $m=i+1,...,d-1$  and $\lambda^{d}_i=-i\sqrt{\frac{2}{i(i+1)}}$.
 (C2) When $i\in\Omega_2\cup\Omega_3$,  the  eigenvalues are $\lambda^1_i=1$, $\lambda^{m}_i=0$ for $m=2,...,d-1$ and $\lambda^d_i=-1.$ It has been shown that for two hermitian $n\times n$  matrices $Q$ and $P$, with eigenvalues
$q_1\geq q_2\geq...\geq q_n$ and $p_1\geq p_2\geq...\geq p_n$, respectively, one has \cite{M},
\begin{equation}\label{QP}
 \sum_{i=1}^{n}q_ip_{n-i+1}\leq Tr(QP)\leq \sum_{i=1}^{n}q_ip_i.
\end{equation}
According to the inequality (\ref{QP}), we have the following three cases.

(D1)  If $i,j\in\Omega_1$,
we have
\begin{equation*}
\Delta(U,i,j)\geq\left\{
\begin{aligned}
&\frac{-2(i+j)}{\sqrt{ij(i+1)(j+1)}},~~~~i+j<d+1,\\
&\frac{-2d}{\sqrt{ij(i+1)(j+1)}},~~~~~~i+j\geq d+1,
\end{aligned}
\right.
\end{equation*}
and
\begin{equation*}
\Delta(U,i,j)\leq \frac{2\left(ij+\min\{i,j\}\right)}{\sqrt{ij(i+1)(j+1)}}.
\end{equation*}
For all $i,j\in\Omega_1$, since $ \frac{2\left(ij+\min\{i,j\}\right)}{\sqrt{ij(i+1)(j+1)}}\geq \frac{2\left(i+j\right)}{\sqrt{ij(i+1)(j+1)}}$, we have
\begin{equation*}\label{D1}
|\Delta(U,i,j)|\leq \frac{2\left(ij+\min\{i,j\}\right)}{\sqrt{ij(i+1)(j+1)}};
 \end{equation*}
 (D2) If $i\in\Omega_1$ and $j\in\Omega_2\cup\Omega_3$
or $i\in\Omega_2\cup\Omega_3$ and $j\in\Omega_1$,
we have
\begin{equation*}\label{D2}
 |\Delta(U,i,j)|\leq\sqrt{\frac{2(\min\{i,j\}+1)}{\min\{i,j\}}};
 \end{equation*}
(D3) If $i,j\in\Omega_2\cup\Omega_3$, we have $|\Delta(U,i,j)|\leq 2$.
Clearly, $|\Delta(U,i,j)|$ is bounded independent of  $U$ for all $i,j\in\Omega$.
On the other hand, Eq.(\ref{F})  implies that
\begin{equation}\label{uuu}
F(\rho)\leq\frac{1}{d^2}+\frac{1}{d^3}\max_{U}\sum_{i,j\in\Omega}
|t_{ij}|\left|\Delta(U,i,j)\right|.
\end{equation}
Based on  the three cases (D1), (D2) and (D3), it is direct to show that
the inequality (\ref{uuu}) gives rise to the inequality (\ref{FUpp}).
$\Box$

Theorem \ref{TH1} immediately provides an upper bound on
FEF in terms of the upper bounds of  $|\Delta(U,i,j)|$. From the fact that  $2|Tr(CD)|\leq Tr(C^2)+Tr(D^2)$
for any $d\times d$ Hermite matrices $C$ and $D$, for any $i,j\in\Omega$, we have
\begin{eqnarray}\label{U2}
|\Delta(U,i,j)|&\leq&\frac{1}{2}\left(Tr\left((U\lambda_iU^{\dagger})^2)\right)+Tr(\lambda^2_j)\right)\\\nonumber
&=&\frac{1}{2}(Tr(\lambda^2_{i})+Tr(\lambda^2_j))=2.
\end{eqnarray}
From inequalities (\ref{uuu}) and  (\ref{U2}) we have the following corollary.

\begin{corollary}\label{c1}
For the quantum state $\rho\in H_A\otimes H_B$,
\begin{eqnarray*}\label{FS1}
F(\rho)&\leq&\frac{1}{d^2}+\frac{2}{d^3}\sum_{i,j\in\Omega}
|t_{ij}|.
\end{eqnarray*}
\end{corollary}

\noindent{\it Example 1}
Consider the following $3\otimes 3$ state \cite{huang},
\begin{eqnarray*}
\rho_{a}&=&\begin{pmatrix}
a&0&0&0&a&0&0&0&a\\
0&a&0&0&0&0&0&0&0\\
0&0&a&0&0&0&0&0&0\\
0&0&0&a&0&0&0&0&0\\
a&0&0&0&a&0&0&0&a\\
0&0&0&0&0&a&0&0&0\\
0&0&0&0&0&0&\frac{1+a}{2}&0&\frac{\sqrt{1-a^2}}{2}\\
0&0&0&0&0&0&0&a&0\\
a&0&0&0&a&0&\frac{\sqrt{1-a^2}}{2}&0&\frac{1+a}{2}\\
\end{pmatrix}.
\end{eqnarray*}
From Theorem \ref{TH1} and Corollary \ref{c1}, we obtain
\begin{equation*}
F(\rho_{a})\leq\frac{3+33a+2\sqrt{1-a^2}}{12+96a}
\end{equation*}
and
\begin{equation*}
F(\rho_{a})\leq\frac{3+\sqrt{3}+(51-\sqrt{3})a+2\sqrt{3(1-a^2)}}{18(8a+1)},
\end{equation*}
respectively. From Fig.1 we see that the upper bound  of the FEF in Theorem \ref{TH1}
is less than the upper bound of Ref.\cite{huang} for $0\leq a\leq0.1849$.
And the upper bound of the FEF in  Corollary \ref{c1} is less than the upper bound of Ref.\cite{huang}  for $0\leq a\leq0.125$.
\begin{figure}[htpb]
\renewcommand{\figurename}{Fig.}
\centering
\includegraphics[width=7.5cm]{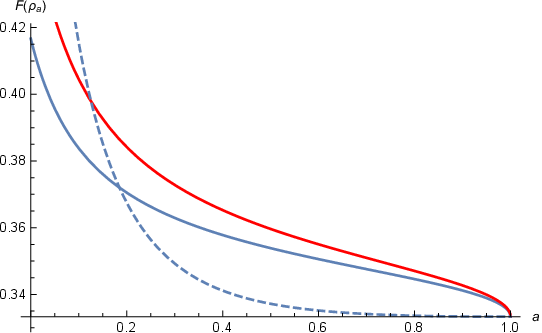}
\caption{{\small  Upper bounds of $F(\rho_{a})$. The blue solid line is from Theorem \ref{TH1}, the red solid line is given by Corollary \ref{c1}, and the dashed line is obtained by Ref.\cite{huang}.}}
\end{figure}

\begin{theorem}\label{TH2}
For any quantum state $\rho\in H_A\otimes H_B$, the singlet fraction $f(\rho)=\langle\phi_{+}|\rho |\phi_{+}\rangle$ has the following expression,
\begin{equation*}\label{FS2}
f(\rho)=\frac{1}{d^2}+\frac{2}{d^3}\left(\sum_{i\in\Omega_1\cup\Omega_2}
t_{ii}-\sum_{i\in\Omega_3}t_{ii}\right).
\end{equation*}
\end{theorem}

{\sf [Proof]}~
Based on Eqs.(\ref{FEF}) and (\ref{F}),  one has
 \begin{eqnarray*}
f(\rho)&=&\frac{1}{d^2}+\frac{1}{d^3}\left(\sum_{i\in\Omega,j\in\Omega_1\cup\Omega_2}
t_{ij}Tr(\lambda_{i}\lambda_j)-\sum_{i\in\Omega,j\in\Omega_3}
t_{ij}Tr(\lambda_{i}\lambda_j)\right)\\\nonumber
&=&\frac{1}{d^2}+\frac{2}{d^3}\left(\sum_{i\in\Omega_1\cup\Omega_2}
t_{ii}-\sum_{i\in\Omega_3}
t_{ii}\right),
\end{eqnarray*}
where the first equality is obtained by taking $U=I$ in Eq.(\ref{F}) and using
$Tr(\lambda_{i}\lambda_j)=2\delta_{ij}$.
$\Box$

In fact, the singlet fraction  is a lower bound of FEF for any quantum states $\rho$, i.e., $F(\rho)\geq f(\rho)$. For the Example 1, we have
$
F(\rho_a)\geq f(\rho_a)=\frac{17a+1}{48a+6}
$ for the state $\rho_a$. In particular, we get $F(\rho_{a=1})=\frac{1}{3}$ from the  upper bound in Theorem \ref{TH1} and $f(\rho)$, see Fig.2.
\begin{figure}[htpb]
\renewcommand{\figurename}{Fig.}
\centering
\includegraphics[width=7.5cm]{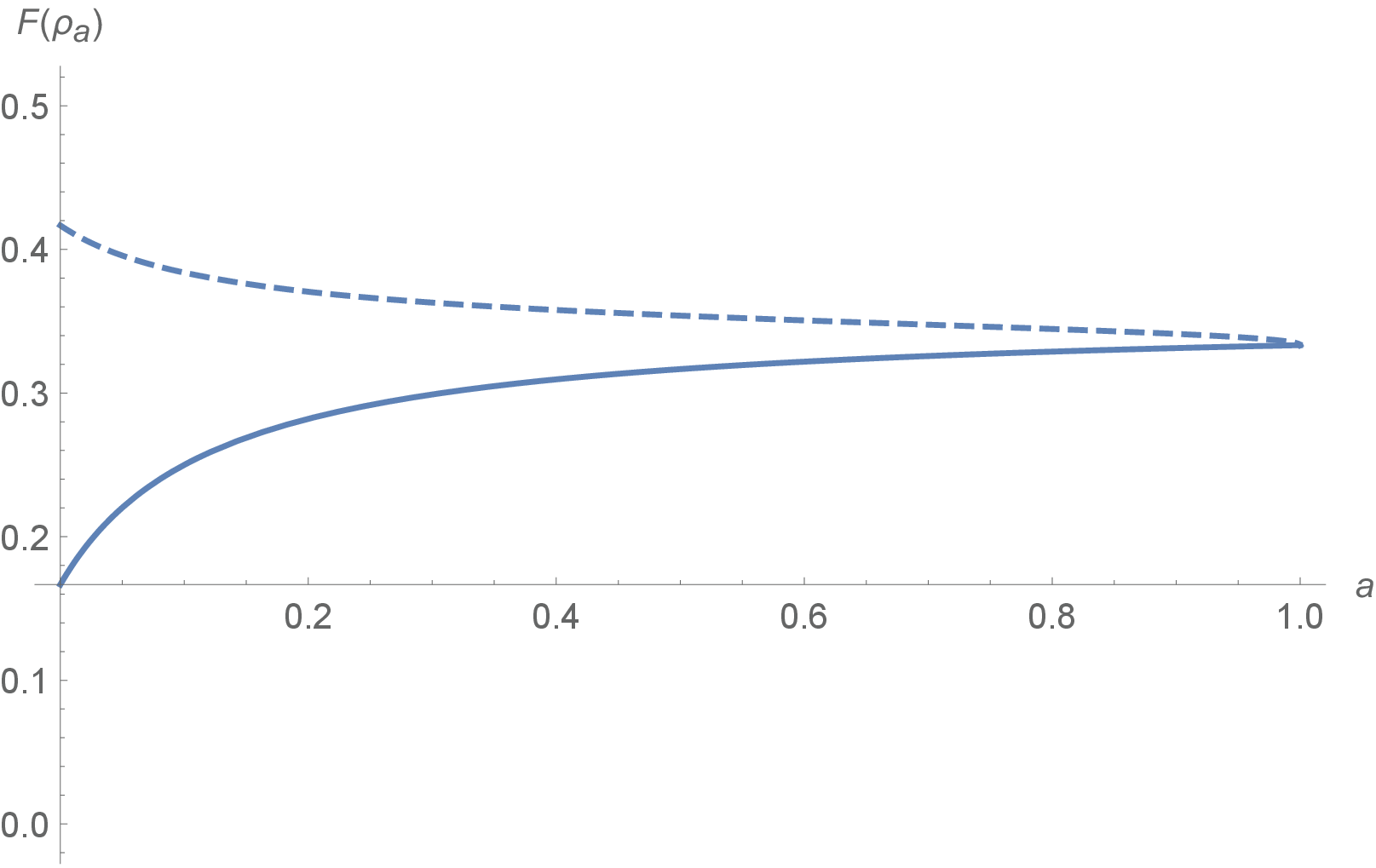}
\caption{{\small  Bounds of $F(\rho_{a})$. The solid line is $f(\rho_a)$,  the dashed line is the upper bound from Theorem \ref{TH1}.}}
\end{figure}

In Ref.\cite{tf}, the authors show that the optimal fidelity $f_{\max}$ of teleportation based on the Bell measurement depends
only on FEF,
\begin{equation*}
f_{\max}(\rho)=\frac{dF(\rho)+1}{d+1}.
\end{equation*}
If $F(\rho)>\frac{1}{d}$, then the state $\rho$ is said to be useful for teleportation \cite{dd4}.

\begin{corollary}\label{c2}
For any quantum state $\rho\in H_A\otimes H_B$, if
\begin{equation*}
\left(\sum_{i\in\Omega_1\cup\Omega_2}
t_{ii}-\sum_{i\in\Omega_3}
t_{ii}\right)>\frac{d(d-1)}{2},
\end{equation*}
then the state $\rho$ is useful  in quantum teleportation.
\end{corollary}

\noindent{\it Example 2}
Consider a $3\otimes 3$ state \cite{up},
\begin{equation*}
\rho=\frac{8}{9}\sigma+\frac{1}{9}|\phi_{+}\rangle\langle\phi_{+}|,
\end{equation*}
where $\sigma=[x|0\rangle\langle0|+(1-x)|1\rangle\langle1|]\otimes [x|0\rangle\langle0|+(1-x)|1\rangle\langle1|].$
According to Corollary  \ref{c2}, we get that when $x\in(0,\frac{2-\sqrt{2}}{4})\cup(\frac{2+\sqrt{2}}{4},1)$,
 $f(\rho)>\frac{1}{3}$ and $\rho$ is useful in quantum teleportation.

Theorem \ref{TH2} also gives rise to an interesting application in entanglement distillation of quantum states. It has been shown that a $U\otimes U^{\ast}$ invariant state $\rho$ can be distilled if and only if $f(\rho)>\frac{1}{d}$ Ref.\cite{dis}, where $U^{\ast}$ denotes complex conjugation for $U$. Consider the $U\otimes U^{\ast}$ invariant isotropic states,
 \begin{equation*}
\rho_{\theta}=\frac{1-\theta}{d^2}I\otimes I+\theta|\phi_+\rangle\langle\phi_+|,
\end{equation*}
where $\frac{-1}{d^2-1}\leq\theta\leq1$. Its Bloch representation is given by
\begin{equation*}
\rho_{\theta}=\frac{1}{d^2}(I\otimes I+\frac{\theta d}{2}\sum_{i=1}^{\frac{(d-1)(d+2)}{2}}\lambda_i\otimes \lambda_i
-\frac{\theta d}{2}\sum_{i=\frac{d(d+1)}{2}}^{d^2-1}\lambda_i\otimes \lambda_i).
\end{equation*}
According to the Theorem \ref{TH2}, we have
\begin{equation*}
f(\rho_{\theta})=\frac{1+\theta(d^2-1)}{d^2}.
\end{equation*}
Thus $\rho_\theta$ can be distilled if and only if $\theta>\frac{1}{d+1}$.

\section{Fully entangled fraction for some classes of mixed states}
From Eq.(\ref{F}), we see that the FEF is independent of the real numbers $r_1,,...,r_{d^2-1}$ and $s_1,s_2,...,s_{d^2-1}$. Therefore, for the following states $\rho_{0}$,
\begin{equation*}\label{rs0}
\rho_{0}=\frac{1}{d^2}(I\otimes I+\sum_{i\in\Omega}r_i\lambda_i\otimes I+\sum_{j\in\Omega}s_j I\otimes\lambda_j),
\end{equation*}
we have $F(\rho_{0})=\frac{1}{d^2}$.

On the other hand, combining Corollary \ref{c1} and Theorem \ref{TH2}, for the isotropic states $\rho_{\theta}$ we have $\frac{1+\theta(d^2-1)}{d^2}\leq F(\rho_\theta)\leq\frac{1+|\theta|(d^2-1)}{d^2}$.
Surprisingly, if $\theta\geq0$, the inequalities in both sides become equalities, and one obtains exactly the
FEF \cite{dd4}. Based on Corollary \ref{c1} and Theorem \ref{TH2} we have the following theorem.

\begin{theorem}\label{TH3}
For any quantum state $\rho\in H_A\otimes H_B$, if
 (E1) $t_{ii}\geq0$ for $ i\in\Omega_1\cup\Omega_2$;
 (E2) $t_{ii}\leq0$ for $ i\in\Omega_3$;
 (E3) $t_{ij}=0$ for $ i\not=j$ and $ i,j\in\Omega$, then
\begin{equation}\label{THK}
F(\rho)=\frac{d+2||T||_{KF}}{d^3}.
\end{equation}
\end{theorem}

{\sf [Proof]}
 Corollary \ref{c1} and Theorem \ref{TH2} imply that
\begin{equation*}
\frac{1}{d^2}+\frac{2}{d^3}\left(\sum_{i\in\Omega_1\cup\Omega_2}
t_{ii}-\sum_{i\in\Omega_3}
t_{ii}\right)\leq F(\rho)\leq \frac{1}{d^2}+\frac{2}{d^3}\sum_{i,j\in\Omega}
|t_{ij}|.
\end{equation*}
Since $t_{ij}$ satisfy (E1), (E2) and (E3), both sides of the inequalities above are equalities,
\begin{equation*}
F(\rho)=\frac{1}{d^2}+\frac{2}{d^3}\sum_{i\in\Omega}|t_{ii}|.
\end{equation*}
Under condition (E3) the correlation
matrix becomes diagonal, so that the  Kay-Fan norm reduces to the sum of absolute
diagonal elements, i.e.,  $||T||_{KF}=\sum_{i\in\Omega}|t_{ii}|$,  thus, we have
\begin{equation*}
F(\rho)=\frac{d+2||T||_{KF}}{d^3},
\end{equation*}
which completes the proof.
$\Box$

It is obvious that Theorem \ref{TH3} applies to the isotropic states $\rho_{\theta}$. Next, we derive the FEF for a family of $3\otimes 3$ states $\rho$ satisfying (E1), (E2) and (E3).

\noindent{\it Example 3}
Let us first consider the following $3\otimes 3$ pure states,
\begin{equation}\label{x}
|\varphi\rangle_{x}=\sqrt{x}|00\rangle+\sqrt{x}|11\rangle+\sqrt{1-2 x}|22\rangle,
\end{equation}
where $x\in[0,\frac{1}{2}]$.
Denote  the density matrix $\rho_{|\varphi\rangle_x}=|\varphi\rangle_x\langle\varphi|$. The coefficients $t^{x}_{ij}=\frac{9}{4}Tr(\rho_{|\varphi\rangle_x}\lambda_i\otimes\lambda_j)$ are give by
$t^{x}_{11}=t^{x}_{33}=\frac{9x}{2}$, $t^{x}_{22}=\frac{6-9x}{2}$, $t^{x}_{44}=t^{x}_{55}=\frac{9\sqrt{x(1-2x)}}{2}$, $t^{x}_{66}=-\frac{9x}{2}$, $t^{x}_{77}=t^{x}_{88}=-\frac{9\sqrt{x(1-2x)}}{2}$ and $t^{x}_{ij}=0$ for all $i\not=j$.
Note that $t^{x}_{ij}$ satisfy (E1), (E2) and (E3). We have
\begin{equation*}
F(\rho_{|\varphi\rangle_x})=\frac{1+2x+4\sqrt{x(1-2x)}}{3}.
\end{equation*}

Generalizing our result to mixed states, we consider a family of $3\otimes 3$ mixed states,
\begin{equation}\label{rhoo}
\rho=\sum_{k}p_k|\varphi\rangle_{x_k}\langle\varphi|,
\end{equation}
where $|\varphi\rangle_{x_k}=\sqrt{{x_k}}|00\rangle+\sqrt{{x_k}}|11\rangle+\sqrt{1-2{x_k}}|22\rangle$,
${x_k}\in[0,\frac{1}{2}]$, $0< p_k\leq1$ and $\sum_k p_k=1$.
By straightforward computing, we obtain
$t_{ij}=\frac{9}{4}Tr(\rho\lambda_i\otimes\lambda_j)
=\sum_kp_k\frac{9}{4}Tr(|\varphi\rangle_{x_k}\langle\varphi|\lambda_i\otimes\lambda_j)=\sum_kp_kt^{x_k}_{ij}$.
Thus, we get
$t_{11}=t_{33}=\frac{9}{2}\sum_kp_kx_k$, $t_{22}=\frac{6-9\sum_kp_kx_k}{2}$, $t_{44}=t_{55}=\frac{9}{2}\sum_kp_k\sqrt{x_k(1-2x_k)}$, $t_{66}=-\frac{9}{2}\sum_kp_kx_k$, $t_{77}=t_{88}=-\frac{9}{2}\sum_kp_k\sqrt{x_k(1-2x_k)}$ and $t_{ij}=0$ for all $i\not=j$,  which
satisfy (E1), (E2) and (E3). From Theorem \ref{TH3}, we obtain
\begin{equation}\label{xk}
F(\rho)=\frac{1+2\sum_kp_k(x_k+2\sqrt{x_k(1-2x_k)})}{3}.
\end{equation}

Note that,  for a state $\rho$ of the form (\ref{rhoo}), if there exists an $x_k\not=0$, then we obtain  $F(\rho)>\frac{1}{3}$ based on Eq.(\ref{xk}).
Hence, all states $\rho$ with at least one  $x_k\not=0$  are useful in quantum teleportation.
For example, taking $\rho_3=y|\varphi\rangle_{\frac{1}{3}}\langle\varphi|+\frac{1-y}{2}|\varphi\rangle_{\frac{1}{2}}\langle\varphi|+
\frac{1-y}{2}|\varphi\rangle_{0}\langle\varphi|$, where $y\in[0,1]$, according to (\ref{xk}) we get $F(\rho_3)=\frac{1+y}{2}$.
It is obvious that $F(\rho_3)>\frac{1}{3}$ for all $y\in[0,1]$ and $\rho_3$ is useful for quantum teleportation.

\section{Conclusion}
In summary, we have studied the fully entangled fraction of arbitrary $d\otimes d$ quantum bipartite states based on the Bloch representation of density matrices. We have presented an analytic upper bound of the FEF, which can be used to improve the previous results. Meanwhile, we have also presented a particular analytical formulae of the singlet fraction $f(\rho)$
by taking $U=I$ in our alternative representation of EFE. Furthermore, we have derived the analytical formulae of FEF for several
classes of quantum mixed states that are useful in quantum teleportation. These results complement the previous ones in
this subject and give a new in evaluating the fully entangled fraction.

\bigskip
\noindent{\bf Acknowledgments}\, \,
This work is supported by the National Natural Science Foundation of China under
grant Nos. 12075159, 12171044 and 12301582; the specific research fund of the Innovation Platform for Academicians of Hainan Province under Grant No. YSPTZX202215; the research
award fund for Natural Science Foundation of Shandong Province under Grant No. ZR2024MA068.

\end{document}